\documentclass[11pt]{article}
\usepackage{amsmath}
\usepackage{amssymb}
\usepackage{amsthm}

\def\fpd#1#2{{\displaystyle\frac{\partial #1}{\partial #2}}}

\def\R{\mathbb{R}}
\def\Im#1{{\mbox{Im}}\,{#1}}
\def\barE{\overline{E}}

\def\onehalf{{\textstyle\frac12}}

\def\sode{{\sc sode}}
\def\pde{{\sc pde}}
\def\ode{{\sc ode}}

\begin{document}

\title{A generalization of Szebehely's inverse problem of dynamics}
\author{W.\ Sarlet$^{a,b}$, T.\ Mestdag$^a$ and G.\ Prince$^b$ \\[2mm]
{\small ${}^a$Department of Mathematics, Ghent University }\\
{\small Krijgslaan 281, B-9000 Ghent, Belgium}\\[1mm]
{\small ${}^b$Department of Mathematics and Statistics, La Trobe University}\\
{\small Melbourne, Victoria 3086, Australia}
}

\date{}

\maketitle

\begin{quote}
{\small {\bf Abstract.} The so-called inverse problem of dynamics is
about constructing a potential for a given family of curves. We
observe that there is a more general way of posing the problem by
making use of ideas of another inverse problem, namely the inverse
problem of the calculus of variations. We critically review and
clarify different aspects of the current state of the art of the
problem (mainly restricted to the case of planar curves), and then
develop our more general approach. }
\end{quote}


\section{Introduction}

There are two kinds of inverse problems which have received
significant attention in the literature, both related to Lagrangian
mechanics in some sense, yet seemingly living separate lives. One is
the so-called inverse problem of the calculus of variations, where
the issue is, as far as classical mechanics is concerned, to study
under what circumstances a system of second-order ordinary
differential equations can be derived from a variational principle.
When the equations of motion are given in normal form, this question
amounts to finding a suitable matrix multiplier which in the end
will be the Hessian of the Lagrangian. The other inverse problem is
often referred to as the inverse problem of dynamics and is, roughly
speaking, as follows. Given a family of paths  in configuration
space, find a potential such that the corresponding classical
Lagrangian system for a particle with unit mass admits the given
family as part of its integral curves. This question seems to have
popped up in the area of celestial mechanics, particularly after
Szebehely launched it in the context of satellite observations in a
much cited paper in 1974 \cite{Szebehely}. Without a variational
content, i.e.\ when the idea is to determine general admissible
forces related to a given family of curves, this problem actually
has a much older history, dating back for example to a paper by
Dainelli of 1880 \cite{Dainelli}. It is astonishing that, in the
extensive literature since Szebehely, when the requirement is
imposed that such forces should fit into a Lagrangian description,
nobody ever approached the problem from the point of view of the
general inverse problem of the calculus of variations: it has always
been taken for granted that the equations should be variational
without allowing for an extra multiplier. Expressed differently, it
is assumed that the kinetic energy term of the Lagrangian under
construction is going to come from the standard Euclidean metric.
Our goal is to fill this gap and thus bring these two separate
inverse problems closer together. To be on the safe side, however,
we should say right away that there have been contributions in which
a general Riemannian metric is used for the kinetic energy function
(see e.g.\ \cite{Melis-Piras1984}, \cite{Borghero1987},
\cite{Mertens1981}). But the idea in those papers is that one starts
the formulation of the problem from a preassigned metric $g$, which
is again altogether a different problem than the one we have in
mind, where the metric in fact is part of the unknowns.

We give a brief sketch of the inverse problem of the calculus of
variations for \sode s (second-order differential equations) in the
next section. For a comprehensive survey of the inverse problem of
dynamics in Section~3, we will limit ourselves here mostly to the
case of two equations ($n=2$). The reason is that there is a
fundamental difference between $n=2$ and $n>2$ in this field, as we
will briefly indicate in that section. Our generalization to allow
for more general multipliers is discussed in Section~4. A number of
examples in the subsequent section will illustrate that our
generalization covers a much wider range of admissible potentials
than those considered so far in the literature.

\section{The inverse problem of the calculus of variations for \sode s}

It will be sufficient for our later purposes that we discuss only
the situation for autonomous second-order differential equations.
Let $\Gamma$ be the \sode\ vector field on a tangent bundle,
modelling the general system of differential equations
\begin{equation}
\ddot{x}^i = F^i(x,\dot{x}), \qquad i=1,\ldots, n. \label{diffeqns}
\end{equation}
The inverse problem of the calculus of variations is the search for
a non-singular symmetric multiplier matrix $g_{ij}(x,\dot{x})$ such
that
\begin{equation}
g_{ij} (\ddot x^j - F^j) \equiv \frac{d}{dt} \left(\frac{\partial
L}{
\partial \dot x^i}\right) - \frac{\partial L}{\partial x^i}, \label{multiplier}
\end{equation}
for some Lagrangian function $L(x,\dot x)$. Necessary and sufficient
conditions for this $g$, generally referred to as the {\it Helmholtz
conditions}, are that
\begin{eqnarray}
&& \Gamma(g_{ij}) = g_{ik}\Gamma^k_j+g_{jk}\Gamma^k_i, \qquad
\Gamma^i_j := -{\frac{1}{2}}\fpd{F^i}{\dot
x^j},  \label{Helmholtz1} \\
&& \fpd{g_{ij}}{\dot{x}^k} = \fpd{g_{ik}}{\dot{x}^j},  \label{Helmholtz2} \\
&& g_{ij}\Phi^j_k  = g_{kj}\Phi^j_i, \qquad \Phi^i_j :=
-\fpd{F^i}{x^j} - \Gamma^i_k\Gamma^k_j - \Gamma(\Gamma^i_j).
\label{Helmholtz3}
\end{eqnarray}
There is an extensive literature about this problem. We limit
ourselves to citing only a few sources \cite{S82, MFLMR, AT92,
CSMBP, GM1, APST, olgageoff, Buca}, where the reader might
appreciate the variety of analytical and geometric methods which
have been used in this field.

For now, it is enough that we look at the more restrictive situation
where $F^i=F^i(x)$ in the given coordinate description of $\Gamma$.
It is then plausible that we restrict the search for a multiplier
also to depend on the position variables only. In that case, the
first condition (\ref{Helmholtz1}) forces the $g_{ij}$ actually to
be constant in those coordinates and all that remains is the
simplified condition (\ref{Helmholtz3})
\begin{equation}
g_{ij}\fpd{F^j}{x^k} = g_{kj}\fpd{F^j}{x^i}, \label{phicond}
\end{equation}
which of course is nothing but the integrability requirement
expressing the existence of a potential function $V(x)$, such that
\[
g_{ij}F^j(x) = - \fpd{V}{x^i}.
\]

\section{The inverse problem of dynamics}

When surveying the vast literature in this field, one gets the
impression that after the early papers on the subject, the situation
has often been obscured by contributions in which the authors have
paid insufficient attention to the sound practice of replacing at
each step an original set of conditions by a new set which is
(generically) equivalent to the previous one. This is in particular
true for the confusion sometimes encountered about the role of the
energy function in the story and about the number of partial
differential equations for admissible potentials to be acquitted. It
is therefore not a waste of time that we try to draw a clear picture
of the overall situation before entering into our new
generalization. As expressed before, the problem becomes quite
different for $n>2$, so we will only discuss the case $n=2$ here.

The basic problem is this: given a family of paths in the form
\begin{equation}
f(x,y)=c, \label{f}
\end{equation}
find a potential $V(x,y)$ and a parameterisation such that the
system
\begin{equation}
\ddot{x} = - V_x, \qquad \ddot{y}= - V_y \label{standardsode}
\end{equation}
has integral curves that are paths belonging to this family for
suitable choices of initial conditions. Most cited in this area is
the so-called Szebehely equation \cite{Szebehely}:
\begin{equation}
f_x V_x + f_y V_y + \frac{2(E-V)}{f_x^2 + f_y^2}(2f_xf_yf_{xy} -
f_x^2 f_{yy} - f_y^2 f_{xx}) =0. \label{Sz}
\end{equation}
This is meant to be a first-order \pde\ for $V$, but what is $E$?
Some papers say that $E$, which as usual represents the total energy
of the system, must be given a pre-assigned constant value. Others
say that it must be pre-assigned as some arbitrarily selected
function of $f$. Bozis, one of the main contributors to the field,
has rightly derived a second-order  \pde\ for $V$ \cite{Bozis1984}
which does not contain $E$, but he took the fact that $E=E(f)$ for
granted to do that. It would be impossible to refer here to every
other statement that has been made about such issues. But the reader
can find a long list of references already in the review paper of
Bozis \cite{Bozis1995}, some less cited contributions are referred
to in \cite{Pal-Anisiu1996}.

Without claiming originality in this section, we start our
comprehensive overview of the situation by some general
considerations. Let $Z$ be a vector field on $\R^2$ whose flow
preserves the given function $f$, meaning that $Z(f)=0$. So $Z$ at
each point of $f(x,y)=c$ represents the tangent direction to a curve
of the family. As is well known, the complete lift of $Z$ to the
tangent bundle $T\R^2$, i.e.\ the vector field
\[
Z^c = Z^i(x)\fpd{}{x^i} + u^j\fpd{Z^i}{x^j}\fpd{}{u^i}
\]
is defined by the property that its flow is the tangent map of the
flow of $Z$ (see e.g.\ \cite{CP}). It follows that the lifts of
integral curves of $Z$ are integral curves of $Z^c$. The idea is
that we want to construct a dynamical system, i.e.\ a \sode\ vector
field
\[
\Gamma = u^i\fpd{}{x^i} + F^i\fpd{}{u^i},
\]
such that $Z^c$ and $\Gamma$ coincide on points in the image of $Z$.
This will guarantee that the lifted integral curves of $Z$ belong to
the set of integral curves of $\Gamma$ (which has the property that
all its integral curves are lifted curves). The condition for that
to happen is that
\begin{equation}
Z^c|_{\Im{Z}} = \Gamma|_{\Im{Z}} \quad \Longleftrightarrow \quad F^i
= Z^j\fpd{Z^i}{x^j}=Z(Z^i). \label{ZGamma}
\end{equation}
Since $Z(Z^i)$ is a function of the $x^i$ only, it suffices to look
for forces $F^i$ which have the same property in the given
coordinates. But more general forces could be allowed provided we
write $F^i|_{\Im{Z}}$ in (\ref{ZGamma}).

A vector field $Z$ for which $Z(f)=0$ is determined up to an
arbitrary factor $h(x,y)$ corresponding to re-parametrization, say
\begin{equation}
Z= h\,Z_0 = h\Big(f_y\fpd{}{x} - f_x\fpd{}{y}\Big). \label{Z}
\end{equation}
The extra condition which comes in now is that we want the \sode\
$\Gamma$ to correspond to a conservative system with potential $V$
and standard (Euclidian) kinetic energy. In other words, we require
that
\begin{eqnarray}
Z(Z^1) &=& Z(h)f_y + h Z(f_y) = - V_x, \label{Vx} \\
Z(Z^2) &=& -Z(h)f_x - h Z(f_x) = - V_y. \label{Vy}
\end{eqnarray}
The essential picture then clearly reads as follows: if $Z(h)$ is
eliminated between these two equations, we obtain an algebraic
relation for $h^2$, which when substituted back in any of the above
equations produces a second-order  \pde\ for $V$. Whenever a
solution for $V$ of this  \pde\ is obtained, $h^2$ will be
determined and we have an admissible \sode\ $\Gamma$. So far, there
seems to be no relation with the energy function or with Szebehely's
equation. However, it is more appropriate to execute the above
process in a slightly different and more efficient way. Eliminating
$Z(h)$ is like multiplying (\ref{Vx}) with $-Z(y)$ and adding the
product of (\ref{Vy}) with $Z(x)$. A system which is algebraically
equivalent with (\ref{Vx}, \ref{Vy}) then follows if we also
multiply (\ref{Vx}) with $Z(x)$ and take the sum with (\ref{Vy})
multiplied by $Z(y)$. The resulting equivalent system reads
\begin{eqnarray}
h^2(f_xZ_0(f_y) - f_yZ_0(f_x)) &=& - (f_xV_x + f_yV_y), \label{1} \\[1mm]
Z\big(\onehalf h^2(f_x^2 + f_y^2) + V\big) &=& 0. \label{2}
\end{eqnarray}
Again, solving (\ref{1}) for $h^2$ and substituting in (\ref{2})
produces the second-order \pde\ for $V$ generated by Bozis. But
(\ref{2}) indicates the first integration of this  \pde. The energy
$E$ is of course a function on $T\R^2$, but we need its restriction
\begin{equation}
\barE := E|_{\Im{Z}} = \onehalf h^2(f_x^2 + f_y^2) + V. \label{barE}
\end{equation}
Then (\ref{2}) implies that $\barE$ must be a function of $f$.
Expressed differently, a first integration of the second-order
\pde\ for $V$ yields the first-order Szebehely equation (\ref{Sz})
in which the term $E$ must be interpreted as an arbitrary function
$\barE(f)$. Integrating this linear first-order  \pde\ will then
introduce a second arbitrary function, $\sigma(\tilde{f})$ say. In
fact, it is clear from the first-order terms in (\ref{Sz}) that the
function $\tilde{f}$ is such that $\tilde{Z}(\tilde{f})=0$, where
$\tilde{Z}$ is any vector field of the form
\begin{equation}
\tilde{Z}= \tilde{h}\,\tilde{Z}_0 = \tilde{h}\Big(f_x\fpd{}{x} +
f_y\fpd{}{y}\Big). \label{tildeZ}
\end{equation}
Note that $Z$ and $\tilde{Z}$ are orthogonal with respect to the
standard Euclidean metric, i.e.\ we have $Z^1\tilde{Z}^1 +
Z^2\tilde{Z}^2=0$. One can further exploit the freedom in $h$ and
$\tilde{h}$ to make them commute. This explains why Broucke and Lass
(\cite{BL}) were able to obtain an expression for the general
solution for $V$ in terms of orthogonal coordinates (see also
\cite{Molnar}, \cite{Puel1995}).

A few comments are in order here concerning the case of higher
dimension $n$. A given family of curves can be specified by $n-1$
relations of the form $f_a(x)=c_a$. The arguments about a dynamical
system $\Gamma$ satisfying a condition of the form (\ref{ZGamma})
remain the same and give rise to a set of $n$ conditions such as
(\ref{Vx},\,\ref{Vy}). This time, elimination of $Z(h)$ gives rise
to $n-1$ algebraic expressions for $h^2$, supplemented by a single
extra condition such as (\ref{2}). Compatibility between the $n-1$
relations for $h^2$, however, creates by itself $n-2$ first-order
{\pde}s for $V$, which is an entirely different story. We will
discuss the case $n>2$ in a subsequent paper. As an aside, there is
also a potentially interesting link with a form of generalized
Hamilton-Jacobi theory. Indeed, the reasoning which led us to
condition (\ref{ZGamma}), if abstraction is made of a given family
of curves and $Z$ therefore is any given vector field on the base
manifold, is precisely what characterizes solutions of the {\sl generalized
Lagrangian Hamilton-Jacobi problem\/} as introduced in
\cite{pepinetal2006}. For an early discussion of a link with the
Hamilton-Jacobi problem, see \cite{Puel1989}.

To complete this section, we still have to report on a slightly
different approach to the same problem, which is more adapted to a
particular type of application. Going back to the requirement
(\ref{ZGamma}), when we first abstain from further restrictions on
the forces and so allow for non-conservative forces with components
$X$ and $Y$, these are simply given by the left-hand sides of
(\ref{Vx},\,\ref{Vy}). In some papers, a first-order  \pde\ for $X$
and $Y$ is set up to study `admissible non-conservative forces'
(see e.g.\ \cite{Bozis1983} and \cite{Anisiu2004b}). For the sake of completeness, we will come back to this point at the end of this section. For now, however, setting up such a \pde\ seems an
unnecessary complication because the left-hand sides of
(\ref{Vx},\,\ref{Vy}) simply give the general expression of such
admissible forces in terms of an arbitrary function $h$. In fact,
these are more or less the expressions discussed  in
\cite{Dainelli}, but it looks better to write them as
functions of say $\eta=h^2$  and its derivatives. Explicitly then they
read,
\begin{eqnarray}
2X &=& \eta_x f_y^2 - \eta_y f_xf_y + 2\eta(f_yf_{xy}-f_xf_{yy}), \label{X} \\
2Y &=& - \eta_x f_xf_y + \eta_y f_x^2 + 2\eta(f_xf_{xy}-f_yf_{xx}).
\label{Y}
\end{eqnarray}
We can subsequently impose the requirement that the forces derive
from a potential, i.e.\ should be of gradient type. The necessary
and sufficient condition
\begin{equation}
X_y = Y_x, \label{XyYx}
\end{equation}
now leads to a second-order  \pde\ for $\eta$ and every solution
will directly lead to an admissible potential by quadratures. That
this version of the problem is equivalent to the one for obtaining
$V$ through Szebehely's equation was demonstrated in \cite{GGP1984},
at least taking for granted that we already know that $\barE$ will
be some function of $f$. We will give a somewhat different
interpretation to a calculation which can be found in that context
in \cite{GGP1984}. Considering (\ref{Sz}) for $V$, with $E$ replaced
by $\barE(f)$, it is natural to replace $V$ by  the function
$2(\barE - V)/(f_x^2+f_y^2)$. This does indeed simplify the equation
for $V$. But this new dependent variable is after all $\eta=h^2$.
The transformed equation is
\begin{equation}
f_x\eta_x + f_y \eta_y + 2\eta (f_{xx}+f_{yy}) = 2
{\barE}^\prime(f). \label{Szforeta}
\end{equation}
This can only be an integrated form of the second-order  \pde\ for
$\eta$ resulting from (\ref{XyYx}), an integrated form which again
has introduced an arbitrary function of $f$. In other words, if we
act with $Z$ or $Z_0$ on (\ref{Szforeta}), the resulting
second-order  \pde\ will be the same as the one following from
(\ref{XyYx}).

Finally, we touch upon an important point which is rarely explained
in the vast literature on the subject. By construction (see
(\ref{ZGamma})) it is only on $\Im{Z}$ that our dynamical system
will have integral curves which are lifts of curves belonging to the
given family. Moreover, the 2-dimensional submanifold $\Im{Z}$ of
$T\R^2$ lies on some 3-dimensional hypersurface of constant energy.
So not all solutions of the \sode\ we constructed will, by far,
correspond to the curves we started from. To pin down solutions
which do have this property, the initial conditions
$(x_0,y_0,\dot{x}_0,\dot{y}_0)$ have to be prescribed as
follows. Starting from an arbitrary initial position $(x_0,y_0)$,
$f(x_0,y_0)$ will fix a constant $c_0$. There is then no more
freedom in selecting an initial velocity. Indeed, for every
particular solution for an admissible $V$, $h^2$ will be fixed, by
(\ref{1}) for example, and $\barE$ as defined by (\ref{barE}) will
be a specific function of $f$, whose constant value $E_0$ therefore
is fixed by $c_0$. Hence, the admissible initial velocity
$(\dot{x}_0,\dot{y}_0)$ is completely determined by the requirements
that it must give a tangent direction to $f$ at the point
$(x_0,y_0)$ and satisfy the relation
$\onehalf(\dot{x}_0^2+\dot{y}_0^2) + V(x_0,y_0) = E_0$. This is in contrast with a series of older papers by Kasner (see \cite{Kasner06} in the first place, and related work in \cite{Kasner09} and \cite{Kasner43}), which are only remotely related to the Szebehely problem, but have been referred to in passing in \cite{Bozis1995}, \cite{Anisiu2004} and \cite{Anisiu2004b}.

Kasner \cite{Kasner06} studied extensively the geometry of an orbit $y=y(x)$ which arises from the solution of a Newtonian system such as $\ddot{x}=X(x,y),\ \ddot{y}=Y(x,y)$. In such a set-up clearly there are three initial values which can be assigned arbitrarily, namely $x_0$, $\dot{x}_0$ and $\dot{y}_0$. Kasner shows that such an orbit satisfies a third-order \ode, which reads
\begin{equation}
y'''(Y-y'X) = y''\big(Y_x + y'(Y_y-X_x) - y'^2X_y\big) - 3y''^2X. \label{Kasner}
\end{equation}
The inverse problem that he poses and solves is about pinning down geometrical characteristics of curves which are also sufficient for orbits $y(x)$ to be generated by a Newtonian system. Clearly, this is a different matter. But for the benefit of the review features of this section, we can establish a link between Kasner's equation (\ref{Kasner}) and our approach to Szebehely's problem for the case that the orbit $y(x)$ under consideration is actually one which belongs to a given family $f(x,y)=c$. For that we go back to the equations (\ref{Vx}, \ref{Vy}) and repeat the steps which led to the equivalent system (\ref{1}, \ref{2}) when the right-hand sides are general functions $X(x,y)$ and $Y(x,y)$, not necessarily coming from a potential. The result is
\begin{eqnarray}
h^2(f_xZ_0(f_y) - f_yZ_0(f_x)) &=& f_xX + f_yY, \label{1gen} \\[1mm]
Z_0\big(\onehalf h^2(f_x^2 + f_y^2)\big) &=& f_yX-f_xY. \label{2gen}
\end{eqnarray}
Solving (\ref{1gen}) for $\eta=h^2$ we get
\begin{equation}
f_xX + f_yY = \eta\,(2f_xf_yf_{xy} - f_x^2 f_{yy} - f_y^2 f_{xx}),
\label{XYeta}
\end{equation}
and substitution of this result in (\ref{2gen}) gives
\begin{equation}
Z_0\left(\frac{1}{2}\, \frac{f_x^2 + f_y^2}{2f_xf_yf_{xy} - f_x^2 f_{yy} - f_y^2 f_{xx}} (f_xX+f_yY)\right) = f_yX-f_xY. \label{new2gen}
\end{equation}
This is the first-order \pde\ for $X$ and $Y$ referred to before, which we will give a different interpretation now. Let $y(x)$ be an orbit belonging to the given family, then $f(x,y(x))\equiv c$. A first differentiation determines $y'$ by $f_x + y'f_y=0$ and a further differentiation provides a formula for $y''$. In addition, along the curve $y(x)$, we have that $Z_0=f_y\, d/dx$. This way, it follows from (\ref{new2gen}) that
\begin{equation}
\frac{d}{dx}\left( \frac{1}{2}\, \frac{1+y'^2}{y''}(Y-y'X)\right) = X+y'X. \label{Kasner2}
\end{equation}
This is a third-order \ode\ for $y(x)$ which indeed is Kasner's equation (\ref{Kasner}). It is worthwhile observing that also this equation can be integrated once in the case that the forces derive from a potential. Indeed, in such a case we have that $X+y'Y= - dV/dx$ and (\ref{Kasner2}) implies that
\begin{equation}
(e-V)y'' = \onehalf (1+y'^2)(y'V_x-V_y), \label{Kasner3}
\end{equation}
where $e$ is an integration constant which of course relates again to the energy integral of the system.

\section{The amalgamated inverse problem}

Looking back at the various features of the Szebehely problem as
summarized so far, it seems to us that the process of determining
admissible forces $(X(x,y),Y(x,y))$ for a given family of paths
$f(x,y)=c$ is very straightforward: there is not even a problem
there, as the expressions (\ref{X},\,\ref{Y}) provide a wealth of
possibilities, in terms of an arbitrary function $\eta$.
Restrictions come in when we require that such forces arise from a
variational principle. But as we asked in the introduction, why then
should a Lagrangian for the system be restricted to have a standard
Euclidean kinetic energy? As explained in Section~2, if we allow a
symmetric, non-singular multiplier $g_{ij}(x,y)$, it must be
constant. That still means, however, that the condition (\ref{XyYx})
of the previous section can be relaxed to
\begin{equation}
g_{12}(Y_y - X_x) = g_{22}Y_x - g_{11}X_y. \label{phicond2}
\end{equation}
This creates a more general second-order  \pde\ for $\eta=h^2$,
containing three extra parameters which can make the difference in
certain applications. We will illustrate this with an extensive
example in the next section. Every solution for $\eta$ of
(\ref{phicond2}) will as before directly lead to an admissible
potential by quadratures, because (\ref{phicond2}) simply guarantees
integrability of the system
\begin{eqnarray}
g_{11}X + g_{12}Y &=& - V_x, \label{newVx} \\
g_{12}X + g_{22}Y &=& - V_y, \label{newVy}
\end{eqnarray}
for $V$, leading to a Lagrangian of the form $L=
\frac{1}{2}g_{ij}\dot{x}^i\dot{x}^j - V$. For now, it remains to
explain how the other aspects of the theory discussed in the
previous section generalize here, in particular with respect to the
Szebehely equation. An observation made with the introduction of the
vector field $\tilde{Z}_0$ in (\ref{tildeZ}) indicates how to
proceed. The point is that passing from the original conditions
(\ref{Vx},\,\ref{Vy}) to the equivalent set (\ref{1},\,\ref{2}) can
be seen as coming from the action on $V$ with the two orthogonal
vector fields $\tilde{Z}_0$ and $Z_0$. With the vector field $Z$
still as in (\ref{Z}), the requirements (\ref{newVx},\,\ref{newVy})
can be written in index notation as
\begin{equation}
g_{ij}Z(Z^j) = - V_{x^i}, \qquad \mbox{with} \qquad Z(Z^j)=
hZ_0(h)Z_0^j + h^2Z_0(Z_0^j). \label{Vxi}
\end{equation}
Observe that the term involving $Z(h)=hZ_0(h)$ has coefficient
$g_{ij}Z_0^j$, i.e.\ comes from the contraction of the metric $g$
with the vector field $Z_0$. The process of elimination of $Z(h)$
between the two equations therefore amounts to taking a further
contraction with the orthogonal vector field
\begin{equation}
Z_0^\perp := (g_{22}f_x - g_{12}f_y)\fpd{}{x} +
(g_{11}f_y-g_{12}f_x)\fpd{}{y}, \label{Z0perp}
\end{equation}
defined (up to a factor) by $g(Z_0,Z_0^\perp)=0$. The term involving
$h^2$ on the other hand has coefficient $g_{ij}Z_0(Z_0^j)$. Now
$Z_0(Z_0^j)$ makes perfectly sense as components of a vector field
in the given coordinates. Indeed, since our metric $g$ is constant
in the given coordinates, the connection coefficients of the
Levi-Civita connection $\nabla$ are zero and $Z_0(Z_0^j)$ are the
components of $\nabla_{\!Z_0}Z_0$. Explicitly, we have
\begin{equation}
\nabla_{\!Z_0}Z_0 = Z_0(f_y)\fpd{}{x} - Z_0(f_x)\fpd{}{y}.
\label{nablaZZ}
\end{equation}
The last point to observe is that multiplication of (\ref{Vxi}) with
$Z_0^i$  (and summation) makes the left-hand side combine into
$Z_0(\onehalf h^2 g_{ij}Z_0^iZ_0^j)$. In conclusion, proceeding in
exactly the same way as in the previous section, the equivalent set
of conditions which generalizes (\ref{1},\,\ref{2}) will read now
\begin{eqnarray}
h^2\, g(\nabla_{\!Z_0}Z_0,Z_0^\perp) = - Z_0^\perp (V), \label{new1} \\[1mm]
Z_0(\onehalf h^2 g_{ij}Z_0^iZ_0^j + V) = 0. \label{new2}
\end{eqnarray}
As before, solving (\ref{new1}) for $h^2$ and substituting into
(\ref{new2}) gives rise to a second-order  \pde\ for $V$, this time
involving three extra parameters $g_{ij}$ as yet to be determined.
Again, (\ref{new2}) indicates that the energy function $E:=\onehalf
g_{ij}\dot{x}^i\dot{x}^j +V$ restricted to $\Im{Z}$ is a function of
$f$:
\begin{equation}
E|_{\Im{Z}} = \barE(f). \label{newbarE}
\end{equation}
As a result, the afore-mentioned second-order \pde\ integrates to a
first-order equation which can be written in the form
\begin{equation}
Z_0^\perp(V) + \frac{2(\barE -
V)}{g(Z_0,Z_0)}\,g(\nabla_{\!Z_0}Z_0,Z_0^\perp) = 0. \label{newSz}
\end{equation}
This is the corresponding generalized Szebehely equation.
Explicitly, for comparison with the original Szebehely equation
(\ref{Sz}),
\begin{equation}
Z_0^\perp(V) + \frac{2(\barE - V)(\det g)}{g_{11}f_y^2 -
2g_{12}f_xf_y + g_{22}f_x^2}\, (2f_xf_yf_{xy} - f_x^2 f_{yy} - f_y^2
f_{xx}) =0. \label{newSz2}
\end{equation}
The link between this equation and the second-order equation
(\ref{phicond2}) for $\eta$ also works in the same way as in the
previous section. It suffices to introduce $2(\barE -
V)/g(Z_0,Z_0)$, which by the energy relation is simply our function
$\eta=h^2$, as new dependent variable in (\ref{newSz}). Then
\begin{eqnarray*}
Z_0^\perp(\eta) &=& \frac{2}{g(Z_0,Z_0)}\Big(\barE'(f)Z_0^\perp(f)-Z_0^\perp(V)\Big)
- \frac{2(\barE - V)}{g(Z_0,Z_0)^2}\,2g(\nabla_{\!Z_0^\perp}Z_0,Z_0), \\
&=& \frac{2}{g(Z_0,Z_0)}\Big(\barE'(f)Z_0^\perp(f) +
\eta\,g(\nabla_{\!Z_0}Z_0,Z_0^\perp) - \eta\,
g(\nabla_{\!Z_0^\perp}Z_0,Z_0)\Big),
\end{eqnarray*}
where we used equation (\ref{newSz}) for replacing $Z_0^\perp(V)$ on
the right. It so happens that $g(Z_0,Z_0)=Z_0^\perp(f)$ and we know
that $g(\nabla_{\!Z_0}Z_0,Z_0^\perp)= -
g(\nabla_{\!Z_0}Z_0^\perp,Z_0)$. This way, the transformed equation
can be written as
\begin{equation}
Z_0^\perp(\eta) + 2\eta\,\frac{g(\nabla_{\!Z_0}Z_0^\perp+
\nabla_{\!Z_0^\perp}Z_0,Z_0)}{g(Z_0,Z_0)} = 2\,\barE^\prime(f).
\label{newSzforeta}
\end{equation}
It is further straightforward to compute that
\[
\nabla_{\!Z_0}Z_0^\perp+ \nabla_{\!Z_0^\perp}Z_0 = (g_{22}f_{xx} -
2g_{12}f_{xy} + g_{11}f_{yy})\,Z_0.
\]
It follows that the above equation for $\eta$ takes the fairly
simple form
\begin{equation}
Z_0^\perp(\eta) + 2\eta\,(g_{22}f_{xx} - 2g_{12}f_{xy} +
g_{11}f_{yy}) = 2\,\barE^\prime(f). \label{newSzforeta2}
\end{equation}
As before, since the right-hand side vanishes under the action of
$Z_0$, acting with $Z_0$ on the left will produce exactly the
second-order  \pde\ for $\eta$ which follows from (\ref{phicond2}).

The discussion in the previous section about admissible initial
conditions remains the same here, with of course an adapted
expression for the energy function $E$.

No doubt some of the results reported in this section, in particular
the generalized Szebehely equation, must correspond to a particular
case for constant $g$ in papers where the analysis starts from a
pre-assigned Riemannian metric. Indeed it can be verified that  the
Szebehely type equation for ``a particle describing orbits on a
given surface'' derived in \cite{Mertens1981} reduces identically to
(\ref{newSz2}) when the $g_{ij}$ are taken to be constant.

In the next section, we want to explore to what extent our
generalization opens up new possibilities for admissible potentials.
The best thing to do then is to look at applications in which the
authors have looked for the existence of a potential within a class
of admissible forces, satisfying a particular ansatz. In such a
case, the best starting point is simply the condition
(\ref{phicond2}) on the forces $X$ and $Y$. It is then of interest
to know that there exists a linear relation between $X$, $Y$ and the
function $\eta$, namely the relation (\ref{XYeta}).
Hence, with given $f$ and an ansatz about the form of $X$ and $Y$,
the best strategy to adopt is the following. First use (\ref{XYeta})
to learn about the effect of the ansatz on $\eta$, and substitute
this information back into one of the general expressions
(\ref{X},\,\ref{Y}) for $X$ and $Y$. This will immediately lead to
restrictions on the forces, leading in general to a number of
different case studies. For each of the subcases identified in the
previous step, imposing the inverse problem condition
(\ref{phicond2}) then should lead to further subcases, for which a
non-singular, constant multiplier $(g_{ij})$ exists, with
corresponding potential $V$. Incidentally, recalling that $\eta$
actually stands for $h^2$, the information coming from (\ref{XYeta})
has led people to introduce what they called `family boundary
curves', which essentially determine from the requirement that
$\eta$ should be positive the boundary of domains in $\R^2$ where
admissible forces of a certain type exist (see e.g.\ \cite{Bozis1994}).

\section{Examples}

1.\ We start with the toy example of families of straight lines. It
is quite trivial to verify that infinitely many potentials for such
a family exist already in the classical picture with the standard
Euclidean metric (see e.g. \cite{Bozis-Anisiu2001}), so there is no
real need for a generalization here. But it is instructive to see
how the generalization works anyway. Let us choose the coordinate
axes in such a way that the straight lines are parallel to the
$x$-axis, so that $f(x,y)\equiv y =c$. The general expressions
(\ref{X},\,\ref{Y}) for admissible forces reduce to
\[
2X = \eta_x, \qquad Y=0.
\]
We can then right away impose the inverse problem condition
(\ref{phicond2}), which reads $g_{12}\eta_{xx} = g_{11} \eta_{xy}$
and integrates to
\[
g_{11} \eta_{y} - g_{12}\eta_{x} = \phi(y), \qquad \phi \mbox{
arbitrary}.
\]
This is in fact the equation (\ref{newSzforeta2}) for this case and
it further integrates to
\[
\eta = \sigma(y) + \psi(g_{11}x+ g_{12}y),
\]
with $\sigma$ and $\psi$ arbitrary functions of the indicated
arguments. Now $X$ becomes $X=\onehalf g_{11}\psi^\prime$ and it
readily follows from (\ref{newVx},\,\ref{newVy}) that $V=-\onehalf
g_{11}\psi$. Hence we have a three-parameter family of kinetic
energy functions and all admissible Lagrangians are of the form
\[
L = \onehalf(g_{11}\dot{x}^2 + 2g_{12}\dot{x}\dot{y} +
g_{22}\dot{y}^2) + \onehalf g_{11}\psi(g_{11}x+ g_{12}y).
\]

2.\ For a more instructive example, take
\begin{equation}
f(x,y):= xy^m, \qquad m\neq 0,\ m\neq -1. \label{Bozisex}
\end{equation}
Bozis \cite{Bozis1994} has carried out a comprehensive analysis
about the existence of a potential for this $f$, starting from the
ansatz that the forces should contain only terms which are quadratic
and cubic in $x$ and $y$. He concluded that there are only two
favourable situations then: one in which $m$ can be kept
unspecified, but both force components then only contain two cubic
terms and only the coefficient of $y^2x$ in $X$ can be left
arbitrary; in the other favourable case $X$ and $Y$ have a quadratic
and two cubic terms but $m=2$ (see also \cite{Bozis1995}). The
values $m=0$ and $m=-1$ are excluded to avoid straight lines. Bozis
actually missed a few cases in his analysis, but anyhow, we will see
that within the same category of admissible forces, our
generalization allows for many more favourable situations.

Say we put
\begin{eqnarray*}
X &=& b_1y^3 + b_2y^2x + b_3yx^2 + b_4x^3 + a_1y^2 + a_2 yx + a_3 x^2, \\
Y &=& r_1y^3 + r_2y^2x + r_3yx^2 + r_4x^3 + s_1y^2 + s_2 yx + s_3
x^2.
\end{eqnarray*}
It follows from the relation (\ref{XYeta}) that for the given $f$,
\[
yX + mxY = m(m+1)x y^{2m-1}\,\eta.
\]
For using this information about $\eta$ into the general expression
(\ref{X}) for $X$, it is computationally appropriate to multiply
both sides of (\ref{X}) with $(m+1)y$. The resulting polynomial
$2(m+1)yX$ on the left then has to match the following expression:
\begin{eqnarray*}
\lefteqn{ (m-3)b_1y^4 + 2(m-1)(b_2+mr_1)y^3x + (3m-1)(b_3+mr_2)y^2x^2  } \\
&&  \mbox{} + 4m(b_4+mr_3)yx^3 + m(5m+1)r_4x^4 + (m-2)a_1y^3 \\
&& \mbox{} + (2m-1)(a_2+ms_1)y^2x + 3m(a_3+ms_2)yx^2 + m(4m+1)s_3
x^3.
\end{eqnarray*}
Immediately, from the coefficients of $x^4$ and $x^3$, it is clear
that we must be in one of the following three cases:
\begin{description}
  \item[Case 1:]\ \ $5m+1=0$ and $s_3=0$,
  \item[Case 2:]\ \ $4m+1=0$ and $r_4=0$,
  \item[Case 3:]\ \ $r_4=s_3=0$.
\end{description}
Similar information comes from the other end of the polynomials,
i.e.\ identification of the coefficients of $y^4$ and $y^3$ shows
that we must be in one of the following three cases:
\begin{description}
  \item[Case a:]\ \ $m+5=0$ and $a_1=0$,
  \item[Case b:]\ \ $m+4=0$ and $b_1=0$,
  \item[Case c:]\ \ $a_1=b_1=0$.
\end{description}
In addition, further identifications of coefficients require that
\begin{eqnarray}
2b_2 &=& (m-1)mr_1, \nonumber \\
(3-m)b_3 &=& m(3m-1)r_2, \nonumber \\
(1-m)b_4 &=& 2m^2r_3, \label{abrs} \\
3a_2 &=& m(2m-1)s_1, \nonumber \\
(2-m)a_3 &=& 3m^2s_2. \nonumber
\end{eqnarray}
What follows is an elementary but rather tedious analysis of
possible combinations. For each subcase of admissible forces we can
identify, we subsequently impose the inverse problem condition
(\ref{phicond2}) to see what further restrictions follow from the
requirement of existence of a potential. We will not give details of
all these calculations, but try to summarize the results in an
appendix. By way of example, here comes a brief discussion of
Case~1.

{\bf Case~1} requires Case~c and the further restrictions
(\ref{abrs}) mean that altogether we will have
\[
m=-1/5,\quad s_3=0,\quad  a_1=b_1=0, \quad r_4\ \mbox{so far
arbitrary}
\]
\[
r_1= (25/3)b_2,\quad r_2=10b_3,\quad r_3=15b_4,\quad
s_1=(75/7)a_2,\quad s_2=(55/3)a_3.
\]
Imposing (\ref{phicond2}) yields five more relations, this time
involving the three extra parameters $g_{ij}$. At this point, we
will not engage into an exhaustive analysis of all possible
subcases, but try to look separately at solutions for a diagonal $g$
(not necessarily the unit matrix) and those with a non-diagonal $g$.
With the choice $g_{12}=0$, we are further limited to
$a_2=a_3=b_3=r_4=0$. The remaining free coefficients are $b_2$ and
$b_4$, with corresponding values for $r_1$ and $r_3$, and since $g$
is determined up to an overall factor, we can for example take
$g_{22}=1$ which then fixes $g_{11}=15(b_4/b_2)$. The potential is
\[
V(x,y) = - \frac{25}{12}\,b_2\,y^4 - \frac{15}{2}\,b_4\,x^2y^2 -
\frac{15}{4}\,\frac{b_4^2}{b_2}\,x^4.
\]
With the choice $g_{12}\neq 0$, say $g_{12}=1$ without loss of
generality, there is some more freedom in the coefficients of the
expression for $X$. For example, $b_2$, $a_2$ and $a_3$ can be
chosen arbitrarily; there are then conditions coming from
(\ref{abrs}) which will fix $b_3$, $b_4$ and $r_4$, and also
$g_{11}$ and $g_{22}$ and the potential can be readily computed.

The conclusions which can be drawn from the more complete analysis
in the appendix is that our generalization clearly identifies more
general forces for which a potential exists in relation to the
family $xy^m=c$ with unspecified $m$. In addition, we find specific
solutions also for the following list of special values for $m$ (in
order of appearance in our discussion): $-1/5,\ -1/4,\ -5,\ -4,\
-2/3,\ -1/2,\ -3/2, \ 2,\ 3,\ -2,\ 1,\ 1/2,\ 1/3$. Note that it is
reassuring that these special values come in pairs such as $(-1/5,
-5)$. This is bound to be the case when we realize that the role of
$x$ and $y$ should be interchangeable.

3.\ With a final example we wish to illustrate another benefit which
the extra freedom incorporated in our generalization can offer. The
point is that the form of the given family of curves may be
suggestive for selecting a multiplier $g$ which will facilitate the
computation of a corresponding potential. Consider a family of
conics represented by
\[
f(x,y) = \frac{1}{2}ax^2 + \frac{1}{2}by^2 + kx, \qquad a,b,k\
\mbox{constant}.
\]
We have that
\[
\Delta:= 2f_xf_yf_{xy} - f_x^2 f_{yy} - f_y^2 f_{xx} = - 2ab\,f -
bk^2.
\]
This is a numerator in the coefficient of $V$ in the generalized
Szebehely equation (\ref{newSz2}). But there is also a denominator
in that coefficient and it so happens that if we choose
\[
g_{11}=a, \quad g_{12}=0, \quad g_{22}=b,
\]
we get that
\[
g_{11}f_y^2 - 2g_{12}f_xf_y + g_{22}f_x^2 = - \Delta,
\]
which clearly considerably simplifies $(\ref{newSz2})$. The equation
reduces to
\[
(ax+k)\fpd{V}{x} + ay\fpd{V}{y} = 2a (\barE (f) - V),
\]
where $\barE$ so far is an arbitrary function of $f$. Using the
method of characteristics, we find for the homogeneous part that
$y=c_1(ax+k)$ upon which the remaining characteristic equation can
be written in the form
\[
\frac{dV}{dx} + \frac{2a}{ax+k}\,V = \frac{2a}{ax+k}\,
\barE(\tilde{f}),
\]
where $\tilde{f}(x):=f(x, c_1(ax+k))$. The solution of this equation
is given by
\[
V = \frac{1}{(ax+k)^2}\,[ c_2 + H(x,c_1)], \quad\mbox{with}\quad
H(x,c_1)= \int 2a(ax+k)\,\barE(\tilde{f})\,dx.
\]
It then follows that the general solution for the Szebehely type
equation is given by
\[
V(x,y) = \frac{1}{(ax+k)^2}\,F\Big(\frac{y}{ax+k}\Big) +
\frac{1}{(ax+k)^2}\,H\Big(x,\frac{y}{ax+k}\Big),
\]
where $F$ is an arbitrary function of the indicated argument. To
obtain a more explicit solution, let us make a choice for the
function $\barE$, say $\barE(f):=f$. The function $H(x,c_1)$ can
then readily be computed and we find that in the end
\[
V(x,y) = \frac{1}{(ax+k)^2}\,F\Big(\frac{y}{ax+k}\Big) +
\frac{1}{2}f(x,y) - \frac{k^2}{4a}.
\]
Note that the additive constant in $V$ can of course be omitted, but
that has a similar effect on the corresponding energy function
$\barE$.

We take this opportunity now to test the consistency of the
different approaches discussed in the previous section. The equation
(\ref{newSzforeta2}) for $\eta$ takes the simple form
\[
Z_0^\perp (\eta) + 4ab\,\eta = 2 \barE^\prime(f),
\]
and with our choice for $\barE$ it explicitly reads
\[
(ax+k)\fpd{\eta}{x} + ay\fpd{\eta}{y} = - 4a\,\eta + \frac{2}{b}.
\]
The general solution of this equation is found to be
\[
\eta(x,y) = \frac{1}{(ax+k)^4}\,G\Big(\frac{y}{ax+k}\Big) +
\frac{1}{2ab},
\]
where $G$ again is an arbitrary function, which should however be
related to $F$ in one way or another. We can now proceed as follows.
With this expression for $\eta$, the relations (\ref{X}, \ref{Y})
provide admissible forces which, in accordance with our choice for
the multiplier $g$, should satisfy the relation (cf.\
(\ref{phicond2})) $bY_x - a X_y=0$. It is straightforward to verify
that this is indeed the case and it implies that we should have (in
agreement with (\ref{newVx}, \ref{newVy})):
\[
a\,X = - V_x, \qquad b\,Y = - V_y.
\]
Since we have computed $V$ in a different way first, these equations
should give us the relation between the arbitrary functions $F$ and
$G$. Putting  $z= y/(ax+k)$, this relation is found to be
\[
F(z) = - \frac{1}{2}\,b(1+ab\,z^2)\,G(z).
\]

\section{Conclusions}

We have introduced a new element into an old problem, which
essentially consists of combining ideas of two different but not
completely unrelated inverse problems. We have argued that there is
a clear distinction between the planar situation ($n=2$) discussed
in the present paper and the problem for $n>2$. We will come back to
the case of higher dimension in a forthcoming paper. Where possible,
we have been cautious about introducing basic concepts and using
notations which make sense also for arbitrary dimension (see for
example the very start of the analysis with conditions
(\ref{ZGamma})). It should also be clear, from the way we formulated
equations such as (\ref{newSz}) and (\ref{newSzforeta}), that
geometrical aspects will become increasingly important when we move
to higher dimensions.

\subsubsection*{Acknowledgements} This work is part of the IRSES
project GEOMECH (nr.\ 246981) within the 7th European Community
Framework Programme. W.\ Sarlet and T.\ Mestdag thank the Australian
Mathematical Sciences Institute for its hospitality. G.\ Prince and
W.\ Sarlet further are indebted to Olga Rossi for useful discussions
and for support from the Czech Science Foundation under grant No
201/09/0981. We are indebted to a referee for bringing the work of Kasner to our attention.

\section*{Appendix}

What follows is a survey of the continuation of the analysis about
the curves $xy^m=c$, as initiated in Section~5.

{\bf Case~2} also requires Case~c and with the conditions
(\ref{abrs}), we will have
\[
m=-1/4,\quad r_4=0,\quad  a_1=b_1=0, \quad s_3\ \mbox{so far
arbitrary}
\]
\[
r_1= (32/5)b_2,\quad r_2=(52/7)b_3,\quad r_3=10b_4,\quad
s_1=8a_2,\quad s_2=12a_3.
\]
The condition (\ref{phicond2}) again produces five more
requirements. With $g_{12}=0$, and then $g_{22}=1$ as before, we get
that $a_3=b_3=0$ while $b_4, b_2, a_2$ remain free and finally: $s_3
= 5 (b_4a_2/b_2)$ and $g_{11}= 10(b_4/b_2)$. The potential is
\[
V(x,y) = - \frac{8}{5}\,b_2\,y^4 - 5\,b_4\,x^2y^2 -
\frac{5}{2}\,\frac{b_4^2}{b_2}\,x^4 - \frac{8}{3}\,a_2\,y^3 -
5\,\frac{b_4a_2}{b_2}\,x^2y.
\]
If, on the other hand, we allow for a non-diagonal multiplier, with
$g_{12}=1$, then $b_2, a_2, a_3$ remain arbitrary and the more
exotic restrictions on the other parameters read
\[
b_3=(49/25)(b_2a_3/a_2), \quad b_4 = (98/125)(b_2a_3^2/a_2^2),\quad
s_3= (72/25)(a_3^2/a_2),
\]
with $g_{11}= -(14/5) (a_3/a_2)$ and $g_{22}= (5/4)(a_2/a_3)$. The
reader can verify that the corresponding potential has five quartic
and four cubic terms.

When we step into Case~3, there are clearly three subcases to
consider. In {\bf Case~3a}, the immediate restrictions are
\[
m=-5,\quad r_4=s_3=0,\quad  a_1=0, \quad b_1\ \mbox{so far
arbitrary}
\]
\[
r_1= (1/15)b_2,\quad r_2=(1/10)b_3,\quad r_3=(3/25)b_4,\quad
s_1=(3/55)a_2,\quad s_2=(7/75)a_3.
\]
For the restrictions coming from the search for a potential, if we
first go for a diagonal $g$ again, so $g_{12}=0$ and $g_{22}=1$, we
find that $a_2=a_3=b_3=b_1=0$, $b_2$ and $b_4$ remain arbitrary and
$g_{11}= (3/25)(b_4/b_2)$. The potential is
\[
V(x,y) = - \frac{1}{60}\,b_2\,y^4 - \frac{3}{50}\,b_4\,x^2y^2
-\frac{3}{100}\,\frac{b_4^2}{b_2}\,x^4.
\]
For a non-diagonal $g$, with $g_{12}=1$, the coefficients become
rather exotic again, so we limit ourselves to stating that $b_1$,
$a_2$ and $a_3$ can be freely chosen; $b_2, b_3, b_4$ and also
$g_{11}, g_{22}$ are subsequently fixed.

In {\bf Case 3b}, the restrictions are
\[
m=-4,\quad r_4=s_3=0,\quad  b_1=0, \quad a_1\ \mbox{so far
arbitrary}
\]
\[
r_1= (1/10)b_2,\quad r_2=(7/52)b_3,\quad r_3=(5/32)b_4,\quad
s_1=(1/12)a_2,\quad s_2=(1/8)a_3.
\]
For a diagonal $g$, with $g_{12}=0$ and $g_{22}=1$, we end up with
$a_2=b_3=0$,\ $a_1, a_3, b_2$ can be left free to choose, upon which
$b_4= (2/5)(a_3b_2/a_1)$ and $g_{11}=(1/16)(a_3/a_1)$. The potential
reads
\[
V(x,y) = - \frac{1}{40}\,b_2\,y^4 -
\frac{1}{32}\,\frac{a_3b_2}{a_1}\,y^2x^2 -
\frac{1}{160}\,\frac{a_3^2b_2}{a_1^2}\,x^4 - \frac{1}{16}\,a_3\,y^2x
- \frac{1}{48}\,\frac{a_3^2}{a_1}\,x^3.
\]
With $g_{12}=1$ on the other hand $a_2$ and $b_3$ need not be zero;
the situation then is that $a_1, a_2, b_2$ can be left arbitrary,
resulting in fixed expressions for $a_3, b_3, b_4$ as well as
$g_{11}$ and $g_{22}$. But as before, listing the corresponding
potential is not very instructive in this case because of exotic
coefficients.

Finally, we look at the more interesting {\bf Case 3c} where $m$ is
not immediately fixed. We have $r_4=s_3=a_1=b_1=0$ and from
(\ref{abrs}) generically,
\[
r_1=\frac{2}{m(m-1)}\,b_2,\quad r_2= \frac{3-m}{m(3m-1)}\,b_3,\quad
r_3= \frac{1-m}{2m^2}\,b_4,
\]
\[
s_1=\frac{3}{m(2m-1)}\,a_2, \quad s_2= \frac{2-m}{3m^2}\,a_3.
\]
Again, there are three special values which will have to be
discussed separately (cf.\ \cite{Bozis1995}): $m=1$, $m=1/2$ and
$m=1/3$. We will do so further on. Continuing first with the generic
case and looking at the five extra identifications coming from the
inverse problem condition (\ref{phicond2}), we first consider the
possibility of a diagonal $g$ again, with $g_{12}=0$ and $g_{22}=1$.
It follows that $a_2=b_3=0$ and either $a_3=0$ or $m=2$. This is in
fact the situation described in \cite{Bozis1995}, but since our
diagonal $g$ need not be the unit matrix, we see that even here a
somewhat more general solution appears, namely (with $m$ arbitrary
for the moment) $b_2$ and $b_4$ can both be left arbitrary, it
suffices to adjust the multiplier by taking
\[
g_{11}= \frac{1-m}{2m^2}\,\frac{b_4}{b_2}.
\]
The potential is given by
\[
V(x,y) = - \frac{1}{2}\,\frac{b_2}{m(m-1)}\,y^4 +
\frac{1}{4}\,\frac{(m-1)b_4}{m^2}\,x^2y^2 +
\frac{1}{8}\,\frac{(m-1)b_4^2}{m^2b_2}\,x^4.
\]
When we look for a non-diagonal $g$ with $g_{12}=1$, the five
identifications coming from (\ref{phicond2}) can be written as
follows,
\begin{eqnarray}
a_2g_{11} &=& a_3\,\frac{(3m+2)(2m-1)}{3m^2}, \nonumber \\
b_3g_{11} &=& b_4\,\frac{(3m-1)(2m+1)}{2m^2}, \nonumber \\
a_3g_{22}\,\frac{2-m}{3m^2} &=& a_2\,\frac{(2m+3)(2-m)}{m(2m-1)}, \label{phicond3} \\
b_3g_{22}\,\frac{3-m}{m(3m-1)} &=& b_2\,\frac{(3-m)(2+m)}{m(m-1)}, \nonumber \\
b_4g_{22}\,\frac{1-m}{m^2} - 2b_2g_{11} &=& 6b_3\,
\frac{(1-m)(1+m)}{m(3m-1)}. \nonumber
\end{eqnarray}
This leads to a quite extensive list of subcases to be considered.
In particular, various new special values for $m$ present
themselves, namely $m=-2/3$, $m=-1/2$, $m=-3/2$, $m=2$ again, $m=3$
and $m=-2$. For all these values a $g$ and corresponding potential
can be found; often of course, it will lead also to a number of the
$b_i$ and $a_i$ having to be zero, so that the forces will contain
only a couple of terms. We limit ourselves to further cases in which
$m$ can be left unspecified. No such cases occur when we take either
$g_{11}=0$ or $g_{22}=0$. Insisting that both should be nonzero, if
also $a_2$ and $a_3$ are nonzero, they are determined by the first
and third condition (\ref{phicond3}). To avoid a conflict with the
excluded value $m=-1$, we must have $b_3=0$ and if no specification
of $m$ is permitted, it further follows that $b_4=b_2=0$. In the
end, the $a_i$ and $m$ are still arbitrary, so we are left with only
quadratic
 forces a
 nd a potential which is given by
\[
V(x,y) = -3\,\frac{(2m+3)a_2^2}{(2m-1)^2a_3}\,y^3 - a_3\, x^2y -3\,
\frac{a_2}{m(2m-1)}\,xy^2 -
\frac{1}{9}\,\frac{(2m-1)(3m+2)a_3^2}{m^2a_2}\,x^3.
\]
If on the other hand $a_2=a_3=0$, then $b_3$ cannot be zero for
general $m$ and $g_{11}, g_{22}$ are determined by the second and
fourth of the relations (\ref{phicond3}). The remaining fifth
condition imposes the following relation between the $b_i$,
\[
2m(1-m)b_3^2 + (3m-1)^2b_2b_4 = 0.
\]
The conclusion here is that $m$ is still arbitrary, that we have
forces with only cubic terms this time and the potential is given by
\begin{eqnarray*}
V(x,y)&=& -2\,\frac{m(m+2)b_3^3}{(3m-1)^3b_4^2}\,y^4
-4\,\frac{b_3^2}{(3m-1)^2b_4}\,y^3x \\
&& \mbox{} - \frac{(m^2-m+1)b_3}{m(3m-1)}\,y^2x^2 - b_4\,yx^3 -
\frac{1}{8}\,\frac{(2m+1)(3m-1)b_4^2}{m^2b_3}\,x^4.
\end{eqnarray*}

It remains to discuss, still for Case~3c, the three special values
of $m$ which were distinguished before imposing the inverse problem
condition and for which no potential was found in \cite{Bozis1994}.

For $m=1$, the analysis is straightforward and the conclusions go as
follows. Remembering that $r_4=s_3=a_1=b_1=0$ already, we also have
$b_2=0$ now. For a diagonal $g$ with $g_{22}=1$, we further must
have $b_3=a_2=a_3=0$ and the potential is given by
\[
 V(x,y) = -\frac{1}{4}\,\left(r_1\,y^4 + g_{11}b_4\,x^4\right).
\]
The fact that $g_{11}$ is unspecified here is the result of a
decoupling of the equations (and it reflects the symmetry between
$x$ and $y$ in this case). For a non-diagonal $g$, $a_2$, $a_3$ and
$b_3$ can be left arbitrary, then $b_4=(5/9)(a_3b_3/a_2)$,
$r_1=5(a_2b_3/a_3)$ and with $g_{12}=1$, $g_{11}=(5/3)(a_3/a_2)$ and
$g_{22}= 15 (a_2/a_3)$. The corresponding potential is
\begin{eqnarray*}
V(x,y)&=& - \frac{75}{4}\,\frac{a_2^2b_3}{a_3^2}\,y^4
- 5\,\frac{a_2b_3}{a_3}\,y^3x - \frac{1}{2}\,b_3\,y^2x^2
-\frac{5}{9}\,\frac{a_3b_3}{a_2}\,yx^3 \\
&& \mbox{} -\frac{25}{108}\,\frac{a_3^2b_3}{a_2^2}\,x^4 -
15\,\frac{a_2^2}{a_3}\,y^3 - 3\,a_2\,y^2x - a_3\,yx^2 -
\frac{5}{9}\,\frac{a_3^2}{a_2}\,x^3.
\end{eqnarray*}

Likewise, for $m=1/2$, $a_2=0$ and for a diagonal $g$ with
$g_{22}=1$, also $a_3=b_3=0$ while $g_{11}=b_4/b_2$. The potential
is
\[
V(x,y)= 2b_2\,y^4  - \frac{1}{2}\,b_4\,y^2x^2 -
\frac{1}{4}\,\frac{b_4^2}{b_2}\,x^4 -\frac{1}{3}\,s_1\,y^3 .
\]
For a non-diagonal $g$ on the other hand, $a_3$, $b_2$ and $b_3$ can
be seen as arbitrary, then $b_4= -2 (b_3^2/b_2)$, $s_1 = -
(5/2)(b_2a_3/b_3)$, $g_{11}= -4 (b_3/b_2)$, $g_{22}=
-(5/2)(b_2/b_3)$. The corresponding potential can easily be
computed. We leave the final case $m=1/3$ to the reader.

\end{document}